\documentclass[twocolumn,amsmath,amssymb,showpacs,pre]{revtex4-1}
\usepackage[export]{adjustbox}
\usepackage{subcaption}
\usepackage{mwe}
\usepackage{ragged2e}

\usepackage{amsmath}
\usepackage{helvet}
\usepackage{courier}
\usepackage{amssymb}
\usepackage{amsthm}   
\usepackage[T1]{fontenc}
\usepackage[latin9]{inputenc}
\usepackage{makeidx}         
\usepackage{graphicx}        
\usepackage[bottom]{footmisc}

\usepackage{hyperref}
\hypersetup{colorlinks=true,
  linkcolor=blue,
  anchorcolor=blue,
  citecolor=red,
  urlcolor=brown,
  pdfauthor={Antonio Moro}}

\providecommand{\be}{\begin{equation}}
  \providecommand{\ee}{\end{equation}}
\providecommand{\bea}{\begin{eqnarray}}
  \providecommand{\eea}{\end{eqnarray}}
\providecommand{\beas}{\begin{eqnarray*}}
  \providecommand{\eeas}{\end{eqnarray*}}

\providecommand{\beni}{\begin{equation*}}
  \providecommand{\eeni}{\end{equation*}}

\providecommand{\bw}{\begin{widetext}}
  \providecommand{\ew}{\end{widetext}}

\def\be{\begin{equation}}
\def\ee{\end{equation}}
\def\bfig{\begin{figure}[htb]}
\def\efig{\end{figure}}

\newcommand{\benumerate}{\begin{enumerate}}
\newcommand{\eenumerate}{\end{enumerate}}

\newcommand{\der}[2]{\frac{\partial #1}{\partial #2}}

\newcommand{\av}[1]{\langle #1 \rangle}

\def\f{\frac}

\date{}

\begin{document}

\author{Paolo Lorenzoni$^{\ast}$ and Antonio Moro$^{\ast \ast}$ \\
\footnotesize{$^{\ast}$ Dipartimento di Matematica e Applicazioni, Universit\`a di Milano - Bicocca, Milano, Italy
~\\
$^{\ast \ast}$Department of Mathematics, Physics and Electrical Engineering, Northumbria University Newcastle\\
Newcastle upon Tyne, United Kingdom }}

\title{An exact study of phase transitions in mean field Potts models}

\date{\today}

\begin{abstract}
We construct the exact partition function of the Potts model on a complete graph subject to external fields with linear and {\it nematic} type couplings. The partition function is obtained as a solution to a linear diffusion equation and the free energy, in the thermodynamic limit, follows from its semiclassical limit. Analysis of singularities of the equations of state reveals the occurrence of phase transitions of nematic type at not zero external fields and allows for an interpretation of the phase transitions in terms of shock dynamics in the space of thermodynamics variables. The approach is shown at work in the case of a $q-$state model for $q=3$ but the method generalises to arbitrary $q$.\\


\end{abstract}

\maketitle


Classical spin models provide a universal paradigm for understanding fundamental mechanisms underpinning the occurrence of critical phenomena and cooperative behaviours in large systems -  from condensed matter physics to combinatorics, from neural networks to biochemistry~(see e.g. \cite{B, Amit, Newman, Agliari,Pagnani}). Originally inspired by specific physical instances, as for example the need to model magnetic properties of crystals, due to their simplicity and generality, spin models can effectively be utilised as a representation of a system of $N$ (possibly large) interacting elements based on a set of rules enforced at the microscopic level. The breadth of applications is vast as the specific statistical rules and distributions do not depend on the nature of the physical interaction\cite{B}.


In this letter, we consider the mean field Potts model~\cite{P,Wu} with external fields. A mean field model is by definition a model where the interaction accounts for all pairs of spins $\sigma_{i}$, that is the spins sit at the vertices of a complete graph.
 The Hamiltonian is given by
\begin{equation}
\label{H}
H_N= -\frac{J}{2 N} \sum_{i,j =1}^{N} \delta(\sigma_{i},\sigma_{j}) -\sum_{j=1}^q h_j\,\sum_{i=1}^N\sigma_i^j 
\end{equation}
where $\delta(\sigma_{i},\sigma_{j})$ is the Kronecker delta function such that $\delta(\sigma_{i},\sigma_{j}) = 1$ if $\sigma_{i} = \sigma_{j}$ and $\delta(\sigma_{i},\sigma_{j}) = 0$ otherwise, and the spin admits $q$ possible values $\sigma_{i} \in \{a_{1},a_{2}, \dots a_{q}\}$. In absence of external fields, i.e. $h_{j} = 0$, the Hamiltonian~(\ref{H}) reduces to the standard Potts model~\cite{P}. The case $q=2$ corresponds to the mean field Ising model also known as Curie-Weiss model~\cite{B}.    

The Potts model has attracted a great deal of interest in relation with modelling thermodynamic systems in physics  as well as a wide range of applications  (see for instance \cite{B,Wu} and reference therein). The two-dimensional model on a square lattice with nearest neighbours interaction exhibits a first order phase transition for $q>4$~\cite{B2}. In the mean field approximation the first-order phase transition occurs for $q>2$. Moreover, it has been conjectured that, for $q$ sufficiently large, the mean-field approximation provides an accurate description of the transition in two or higher dimensions~\cite{MS}.

Although several variants of the Potts model with external fields have been extensively studied in the literature, at the best of our knowledge the Hamiltonian of the form \eqref{H} has not been previously considered. In the approach detailed below, we treat the Hamiltonian~\eqref{H} as a deformed ({\it dressed}) version of the model with zero external fields, i.e. $h_i = 0$. We show that the partition function satisfies a linear diffusion equation and in the large $N$ limit, the free energy is consequently obtained as a solution of a Hamiltoni-Jacobi equation, equivalent to the problem of a free particle in $q-1$ dimensions with suitable initial conditions. Importantly, the exact expression for the free energy and the equations of state with external fields provide a novel and simple representation of the solution for the mean field Potts model in terms of the moments. This allows for the exact description of critical sets and phase transitions.  For the sake of simplicity, we focus on the case $q=3$ but the proposed approach naturally extends to arbitrary $q$.


{\it Equations of state.} In order to derive the equations of state, following the approach introduced in \cite{BDGM} and further developed in \cite{BM,DGM,ABSM}, we look for a differential identity satisfied by the partition function
\begin{equation}
\label{ZNgen}
Z_{N} = \sum_{\{ {\cal C}_{N} \}} \; e^{- \beta H}
\end{equation}
where the sum runs over all spin configurations $ {\cal C}_{N}$ and $\beta = 1/T$ where $T$ is the temperature.
The crucial step in the derivation of the required identities is the observation that the above Kronecker's delta function admits the following polynomial representation 
\begin{equation}
\label{kronecker}
\delta(\sigma_i,\sigma_j)=\sum_{l=1}^q\prod_{k\ne l}\f{\sigma_i-a_k}{a_l-a_k}\f{\sigma_j-a_k}{a_l-a_k}.
\end{equation}
 For $q=3$ with $\sigma_{i} \in \{-1,0,1\}$ the Kronecker's delta~\eqref{kronecker} reads as
\begin{equation}
\label{deltaq3}
\delta(\sigma_{i},\sigma_{j}) = \frac{3}{2} \sigma_{i}^{2} \sigma_{j}^{2} + \frac{1}{2} \sigma_{i} \sigma_{j} - \left(\sigma_{i}^{2} + \sigma_{j}^{2} \right) + 1,
\end{equation}
leading to the Hamiltonian of the form
\begin{equation}
H=-\frac{NJ}{2}\left(\frac{1}{2} \mu_{1}^{2} +\frac{3}{2} \mu_{2}^{2}  - 2 \mu_{2} \right) 
- N (h_{1} \mu_{1} + h_{2} \mu_{2}),
\end{equation}
where $\mu_{1} =\sum_{i=1}^N\sigma_i/N$ and $\mu_{2}=\sum_{i=1}^N\sigma_i^2/N$ are the first and second moments. Introducing the re-scaled variables $t=\beta J/2$, $x=\beta h_1$ and $y=\beta h_2$, one can immediately verify that the partition function~\eqref{ZN} reads as
\begin{equation}
\label{ZN}
Z_{N} =\sum_{\{ \mathcal{C}_N\} }e^{N\left[t\left(\f{1}{2} \mu_{1}^2+\f{3}{2} \mu_{2}^2-2\mu_{2} \right)
+x  \mu_{1}+y \mu_{2} \right]},
\end{equation}
and identically satisfies the following linear diffusion equation
\begin{equation}
\label{Zdiff}
Z_{N,t}+2Z_{N,y}=\f{1}{N}\left(\f{1}{2}Z_{N,xx}+\f{3}{2}Z_{N,yy}\right)
\end{equation}
with the notation $Z_{N,t} = \partial Z_{N}/ \partial t$ and so on. The associated initial condition 
\begin{equation}
\label{Zinit}
Z_{N,0}(x,y)= Z_{N}(x,y,0) = (1+2e^y\cosh{x})^N,
\end{equation}
calculated by recursion, is the partition function of a system of non-interacting spins coupled to the constant external fields $x$ and $y$. 

In order to study the behaviour of the system in the thermodynamic limit, i.e. when $N \to \infty$, we introduce the free energy function as 
\begin{equation}
F_{N}=\f{1}{N}\log{Z_{N}}.
\end{equation}
We shall emphasise that {\it physical} free energy is ${\cal F}_{N} = -F_{N}$, so that the equilibrium corresponds to a maximum of $F_{N}$, i.e. a minimum of ${\cal F}_{N}$.   The definition of $F_{N}$ can be viewed as the inverse Madelung transform that, applied to the equation~(\ref{Zdiff}), gives the following Hamilton-Jacobi type equation with diffusion term of order $O(1/N)$
\begin{align}
\label{eqF}
F_{N,t}+2F_{N,y}&=\f{1}{2}F_{N,x}^2+\f{3}{2}F_{N,y}^2+\f{1}{N}\left(\f{1}{2}F_{N,xx}+\f{3}{2}F_{N,yy}\right).
\end{align}
The associated initial condition is
\begin{equation}
\label{CD}
F_{N}(x,y,0)=\log{ \left (1+2 e^{y}\cosh{(x)} \right)}.
\end{equation}
Equations of the type~\eqref{eqF} have been studied in the one-dimensional case from the point of view of integrability in~\cite{ALM} and naturally arise in the constext of mean field models (see e.g. \cite{BM,BDGM}).
Unlike $Z_{N}(x,y,0)$, which diverges for large $N$,  the rescaled variable $F_{N}$ corresponds to an initial datum independent of $N$. 
The above formalism allows to calculate the free energy (and its large $N$ asymptotics) as a a solution of a well-posed initial value problem and derive physical observables and order parameters by differentiation with respect to the conjugated thermodynamic variables.
For instance, the expectation values of the moments $m_{k,N} := \av{\mu_{k}}_{N}$ for $k=1,2$  are given by
\begin{equation}
m_{1,N} = \der{F_{N}}{x}, \qquad m_{2,N} = \der{F_{N}}{y}.
\end{equation}
The free energy function $F(x,y,t)$ in the thermodynamic limit can be obtained, far from singularities, as a solution to the equation~\eqref{eqF} where the diffusion term in neglected, i.e. 
\begin{equation}
\label{HJ}
F_t+2F_y-\f{1}{2}F_x^2-\f{3}{2}F_y^2=0
\end{equation}
with initial condition $F(x,y,0) = F_{N} (x,y,0)$.
The solution of the Hamilton-Jacobi type equation~(\ref{HJ}) yields a free energy of the form
\begin{equation}
\label{Fsol}
F = x m_{1} + y m_{2} + \sum_{k=1}^{3} p_{k}^{2} t  - \sum_{k=1}^{3} p_{k} \log p_{k}
\end{equation}
where the quantities $p_{k}$, $k=1,2,3$ are interpreted as probabilities of observing the spin states $+1,-1,0$, respectively, and parametrised in terms of the moments as follows
\[
p_{1} = \frac{m_{1} + m_{2}}{2} \quad p_{2} = \frac{m_{2} - m_{1}}{2} \quad p_{3} = 1- m_{2}.
\]
Obviously, $\sum_{k=1}^{3} p_{k} = 1$. For $x=y=0$ the expression~(\ref{Fsol}) is consistent with the standard mean field solution (see e.g. \cite{Mussardo}) but the method provides us with the explicit parametrisation of the probabilities $p_{k}$ in terms of the moments.
The moments $m_{1}(x,y,t)$ and $m_{2}(x,y,t)$ play the role of order parameters and are obtained from the equations of state given by the stationary points of the free energy~\eqref{Fsol} 
\begin{gather}
\label{ESq3}
\begin{aligned}
\psi_{1} & := x + m_{1} t  - \frac{1}{2} \log \frac{m_{1} + m_{2}}{m_{2} - m_{1}} = 0 \\
\psi_{1} & := y + (3 m_{2} - 2) t - \frac{1}{2} \log \frac{m_{2}^{2} - m_{1}^{2}}{4 (m_{2} - 1)^{2}} = 0.
\end{aligned}
\end{gather}

Importantly, equations~(\ref{ESq3}) provide closed set of equations for the first two moments. They give the mean field solution of the {\it dressed} Potts model as well as, in the limit of vanishing fields $x,y \to 0$, the mean field solution for the standard model. 

\begin{figure}[htb]
\centering
\begin{subfigure}[b]{0.18\textwidth}
\centering
    \includegraphics[width=\textwidth]{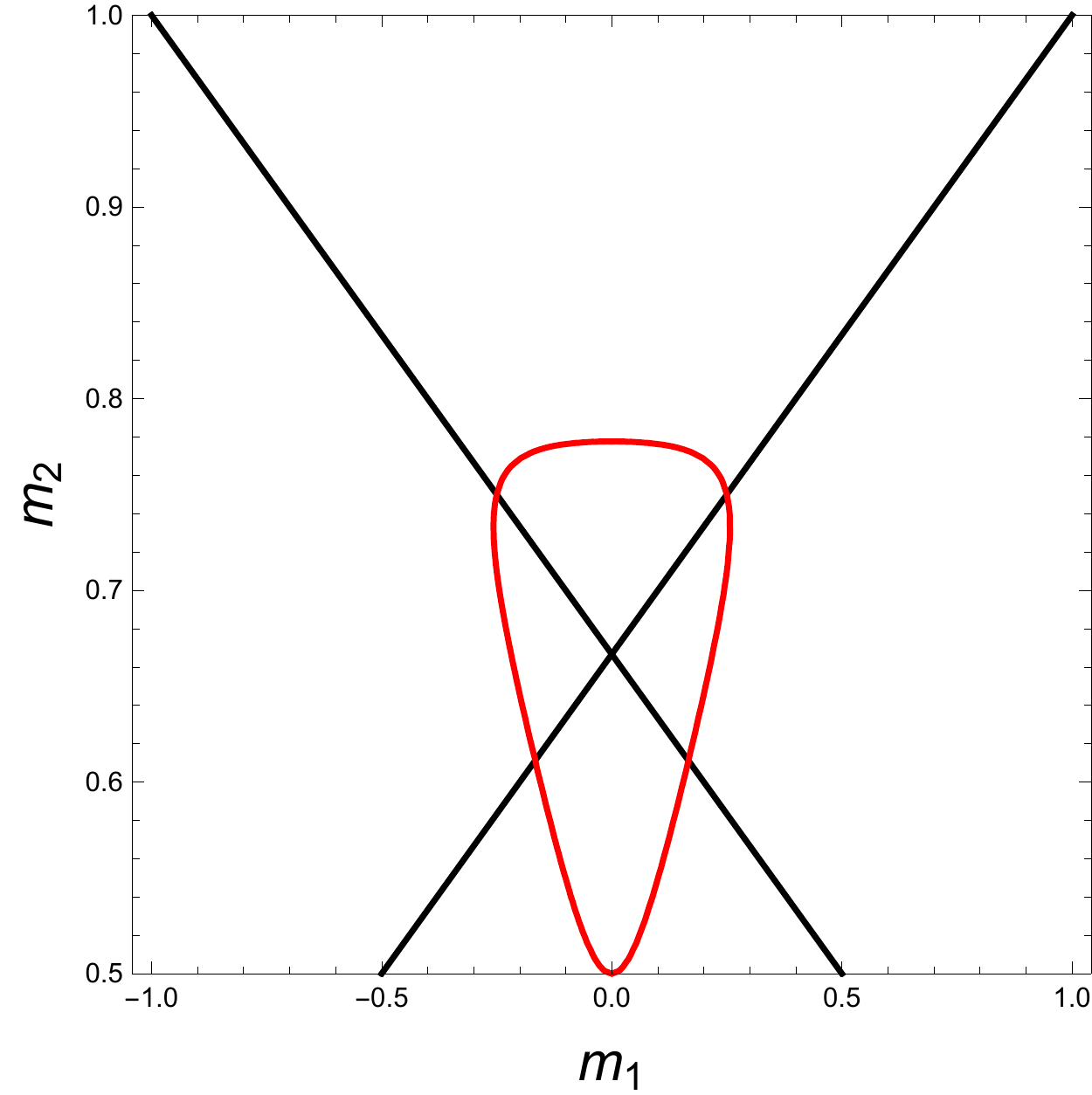} \caption{ }
    \label{fig:critsetm1}
  \end{subfigure}
\quad 
  \begin{subfigure}[b]{0.19\textwidth}
  \centering
    \includegraphics[width=\textwidth]{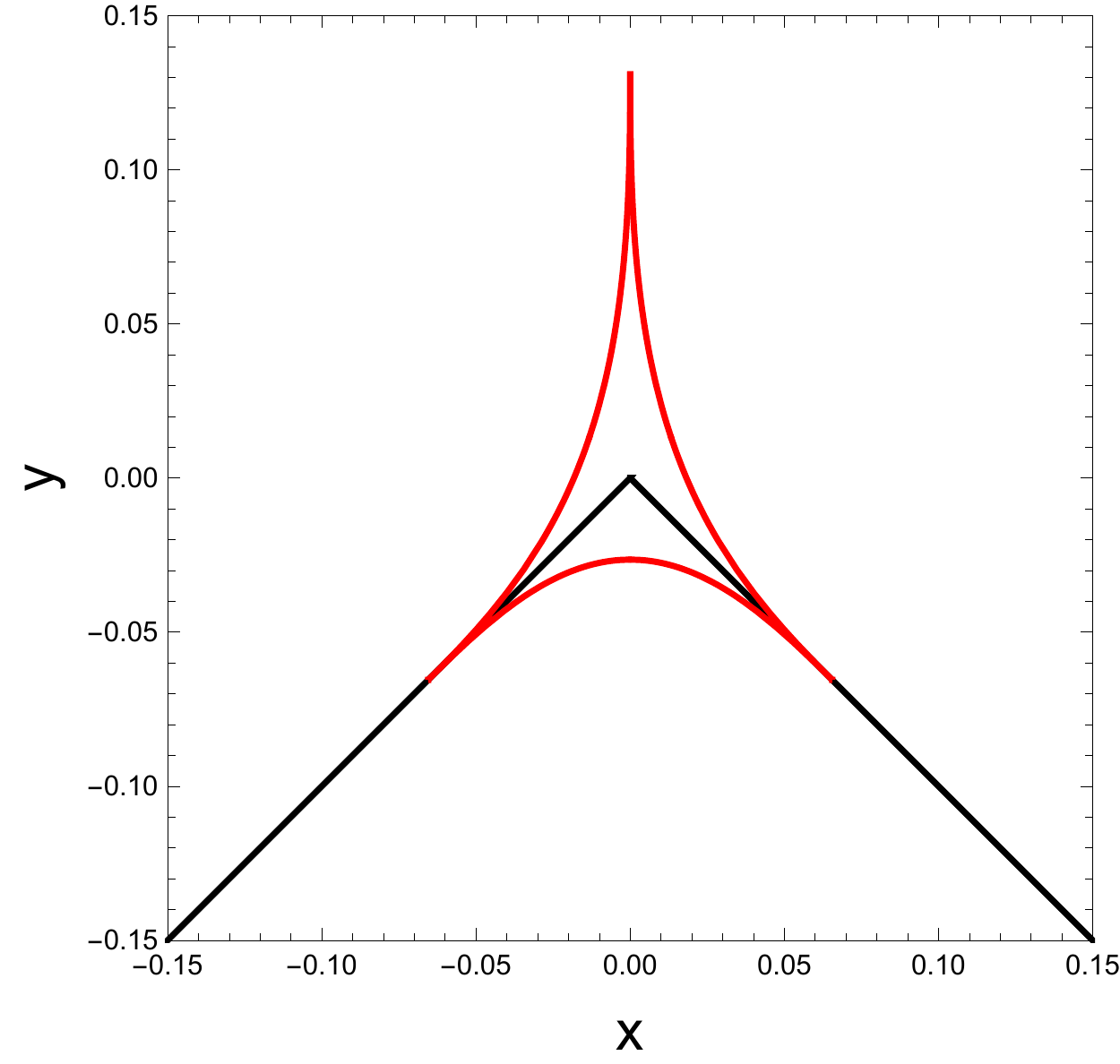}\caption{}
    \label{fig:critsetm2}
  \end{subfigure}\\\vspace{10pt}
\caption{\small Locus of cusp points on the (a) $(m_{1},m_{2})$  and (b) $(x,y)$ plane. }
  \label{fig:critsetm}
\end{figure}

{\it Critical behaviour.} The analysis of the critical sector and singularities of the equations of state for the moments requires the study of the family of maps of the plane induced by the  equations of state~(\ref{ESq3}), i.e.
\begin{align*}
\Psi : \quad (m_{1},m_{2}) \in [-1,1] \times [0,1] \to (\psi_{1},\psi_{2}) \in  \mathbb{R}^{2}.
\end{align*}
Thermodynamic variables $(x,y,t)$ parametrise the family of maps. We note that singularities of maps of the plane are completely classified and they are either folds or cusps~\cite{W}. Cusps are interpreted as the critical points associated to the phase transition of the underlying system.
Introducing the Jacobian $J$ of the map $\Psi$,
cusp points are characterised by the following conditions
\begin{subequations}
\label{cuspcond}
\begin{align}
\label{Jcond}
J := \der{\psi_{1}}{m_{1}}\der{\psi_{2}}{m_{2}}-\der{\psi_{1}}{m_{2}}\der{\psi_{2}}{m_{1}}&= 0  \\
\label{tancond}
\der{\psi_{i}}{m_{2}}\der{J}{m_{1}}-\der{\psi_{i}}{m_{1}}\der{J}{m_{2}}&= 0, \qquad i = 1,2.
\end{align}
\end{subequations}
In particular, the equation~\eqref{Jcond} defines the general fold, i.e. the set where the Jacobian of the map $\Psi$ is singular; conditions~\eqref{tancond} mean that the gradient of the map is tangential to the general fold. We also note that, subject to the condition~(\ref{Jcond}), equations~(\ref{tancond}) are linearly dependent.
Moreover, equations~(\ref{Jcond}) and $\eqref{tancond}$ imply that the locus of cusp points  on the $(m_{1},m_{2})$ plane is given by the union of two straight lines and a quartic curve. Their equations are 
\begin{subequations}
\begin{align}
\label{line_I}
\textup{(I)} \qquad &m_{1} - 3 m_{2} + 2 = 0 \\ 
\label{line_II}
\textup{(II)} \qquad &m_{1} + 3 m_{2} - 2 = 0  \\
\nonumber
\textup{(III)} \qquad & 2m_1^4+18m_2^4+12m_1^2m_2^2-41m_1^2m_2\\
\label{line_III}
&-23m_2^3+25m_1^2+7m_2^2=0
\end{align}
\end{subequations}
and solutions are shown in Figure~\ref{fig:critsetm}. 
Equations of states~\eqref{ESq3} allow to describe the ``dynamics'' of the cusp points in the parameter space $(x,y)$ with respect to the ``time'' $t$. In particular, the equations~(\ref{cuspcond}) imply that {\it critical time} at which a cusp singularity occurs along the lines~(\ref{line_I}) and ~(\ref{line_II}) is
\[
t_{c}^{(I)} \left ( m_{2} \right ) = t_{c}^{(II)} \left ( m_{2} \right ) = \frac{1}{2 (1-m_{2})}  \qquad m_{2} \in  \left [\frac{1}{2},1 \right ]
\]
where $m_{2}$ is used to parametrise the lines in both cases. We note that cusp sets of similar structure arise in the context of nematic liquid crystals models~\cite{DGM}. The minimum critical time at which the cusp first occurs is 
\[
t_{c,min}^{(I),(II)} = t_{c}^{(I)}\left ( \frac{1}{2}\right ) = t_{c}^{(II)}\left ( \frac{1}{2}\right ) = 1.
\]
We note that the cusp exists for $m_{2} \geq 1/2$ as for $m_{2}< 1/2$ the equation of state returns to complex values of $x$ and $y$.
We now study the critical time of cusp points along the loop~(\ref{line_III}). 
The minimum time at which a cusp occurs on the loop is
\[
t_{c,min}^{(III)} = t_{c}^{(III)}\left (\frac{11}{18} \right) = t_{c}^{(III)}\left (\frac{7}{9} \right) = \frac{9}{7}.
\]
The value $m_{2} = 11/18$ gives the bottom intersection between the loop and the straight lines and $m_{2} = 7/9$ corresponds to the upper extreme of the loop as shown in Figures~\ref{fig:critsetm}.
The cusp dynamics can be summarised as follows: the first two cusp points are simultaneously created at the time $t = 1$ at the bottom of the two straight lines and travel upward until they hit the loop at the time $t = 9/7$. At the same time an extra cusp is created at the top of the loop corresponding to $m_{2} = 7/9$. Cusps generated at the intersection of the straight lines and the loop split in three cusps: one continues to propagate along the line and other two propagates along the loop in opposite direction. The cusp generated at the top of the loop splits in two cusps propagating also in opposite directions. Cusps traveling against each other along the loop will collide and annihilate at the time $t = 4/3$ corresponding to the values $m_{2} = 3/4$ (second intersection with the straight lines) and $m_{2} = 1/2$ (lower extreme of the loop).
The locus of cusp points on the $(m_{1},m_{2})$ is mapped onto the $(x,y)$ through the moments equations of state~(\ref{ESq3}). The semi-lines are mapped onto the following semi-lines via the equations
\begin{align*}
&x^{(I)} = y^{(I)} = \frac{2 -3 m_{2}}{2 (1-m_{2})} + \frac{1}{2} \log \frac{2 m_{2} - 1}{1-m_{2}} \\
&x^{(II)} = - y^{(II)} = - \frac{2 -3 m_{2}}{2 (1-m_{2})} - \frac{1}{2} \log \frac{2 m_{2} - 1}{1-m_{2}}, 
\end{align*}
where $m_{2} \in \left [ \frac{1}{2}, 1 \right ] $, and the loop is mapped onto the curve triangular region shown in Figure~\ref{fig:critsetm}(b) via the equations
\begin{align*}
& x^{(III)} =\pm \left ( \frac{2 \beta}{3 (1-m_{2}) (5 - \alpha)} + \frac{1}{2} \log \frac{2 m_{2} -\beta}{2 m_{2} + \beta} \right)  \\
& y^{(III)} =  - \frac{4 (3 m_{2} -2)}{3 (1-m_{2}) (5-\alpha)} + \frac{1}{2} \log \frac{(2 m_{2} - \beta) (2 m_{2} + \beta)}{16 (1 - m_{2})^{2}},
\end{align*}
where $m_{2} \in \left [\frac{1}{2},\frac{7}{9} \right]$ and $\alpha = \sqrt{25 - 32 m_{2}}$,  $\beta = \sqrt{41 m_{2} - 12 m_{2}^{2} - 25 + 5 \alpha (1-m_{2})}$.
\begin{figure}[htb]
\centering
\begin{subfigure}[b]{0.14\textwidth}
    \includegraphics[width=\textwidth]{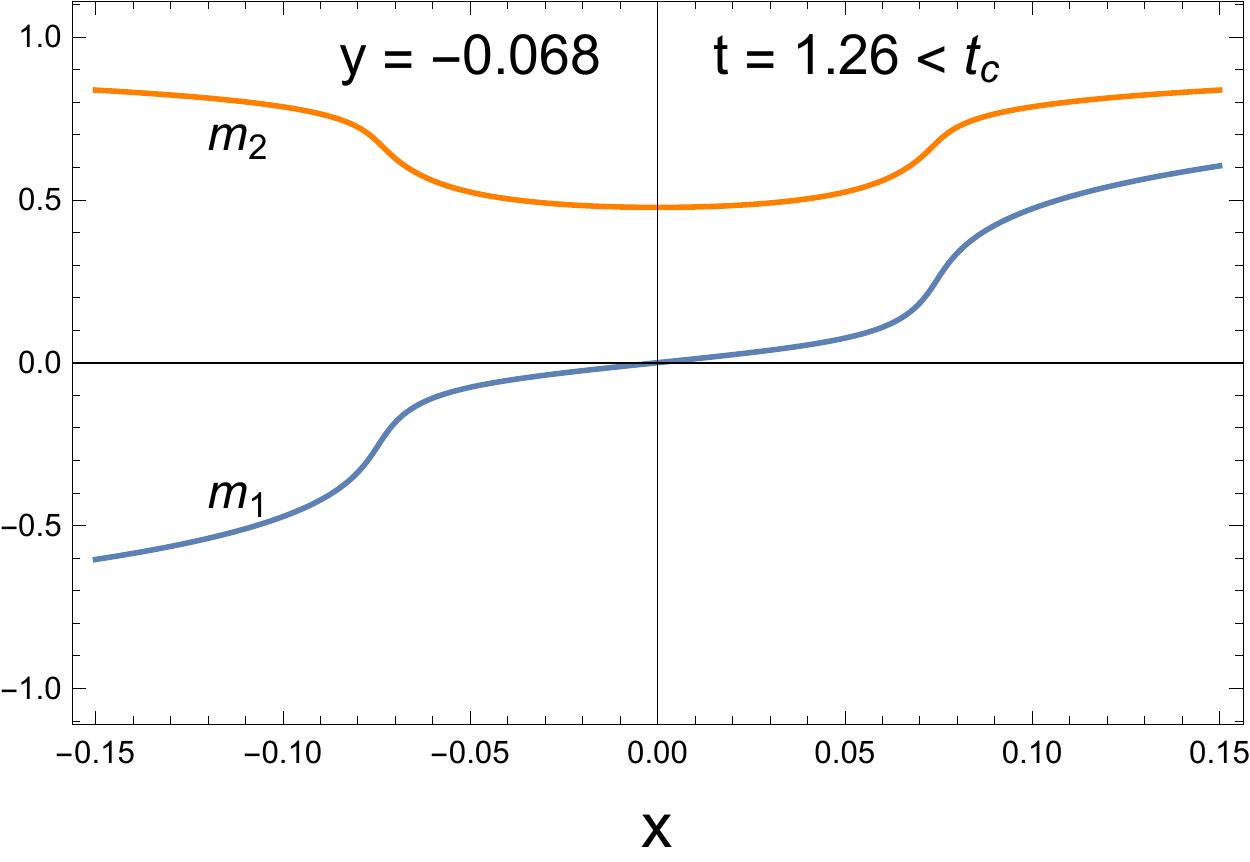}   
  \end{subfigure}
  \quad 
  \begin{subfigure}[b]{0.14\textwidth}
    \includegraphics[width=\textwidth]{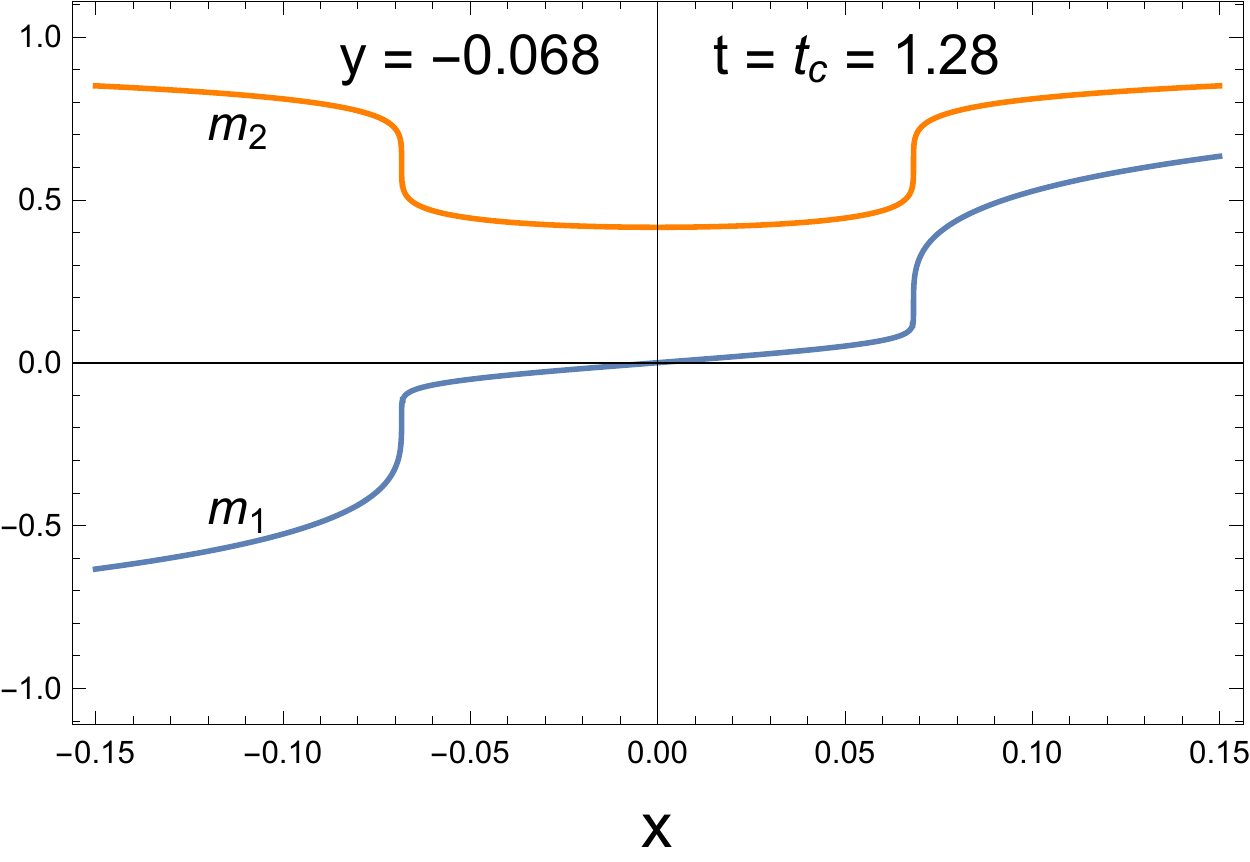}
  \end{subfigure}
  \quad 
  \begin{subfigure}[b]{0.14\textwidth}
    \includegraphics[width=\textwidth]{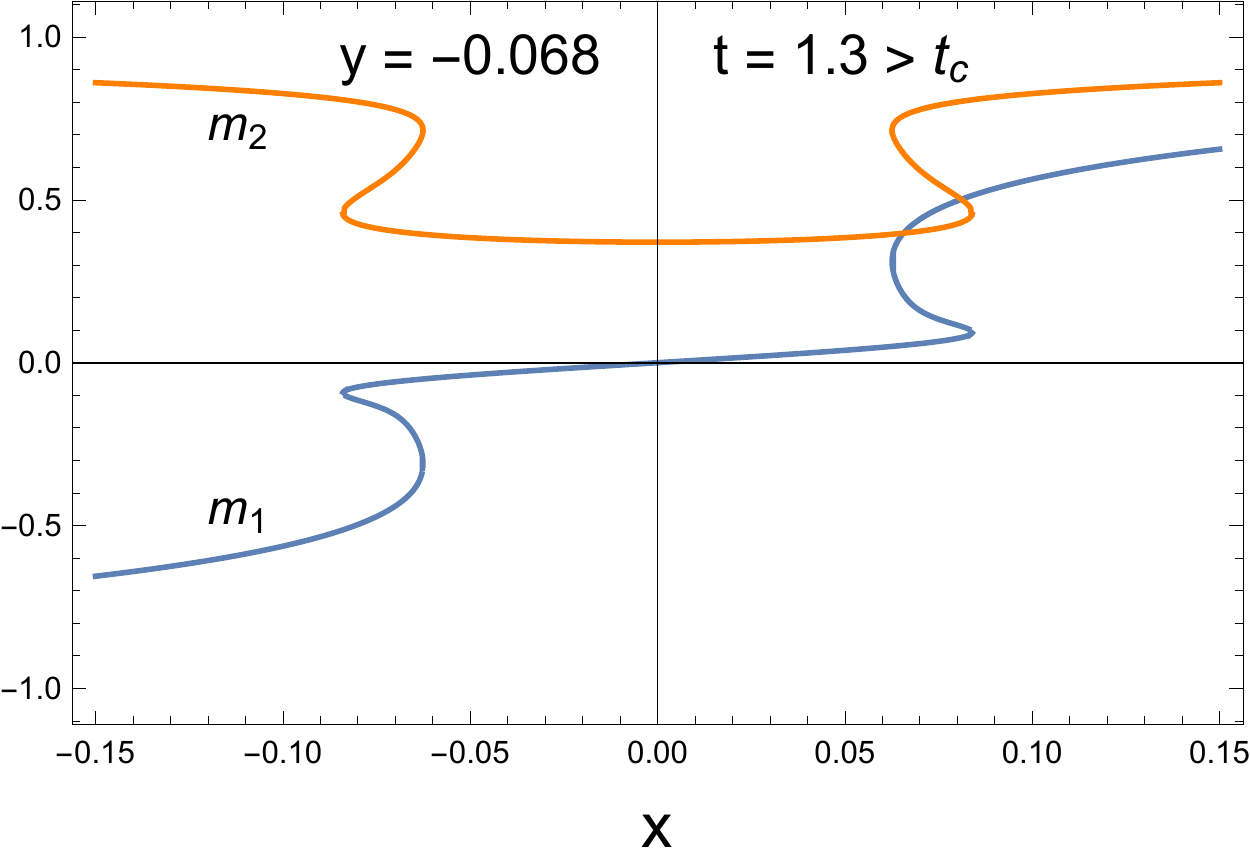}
  \end{subfigure} 
\caption{\small 
Gradient catastrophe and multivaluedness of the order parameters along the line $y \simeq - 0.068$. The critical time is $t_{c} \simeq 1.28$.}
  \label{fig:critprofile}
\end{figure}

\begin{figure}[htb]
\begin{subfigure}[b]{0.25\textwidth}
    \includegraphics[width=\textwidth]{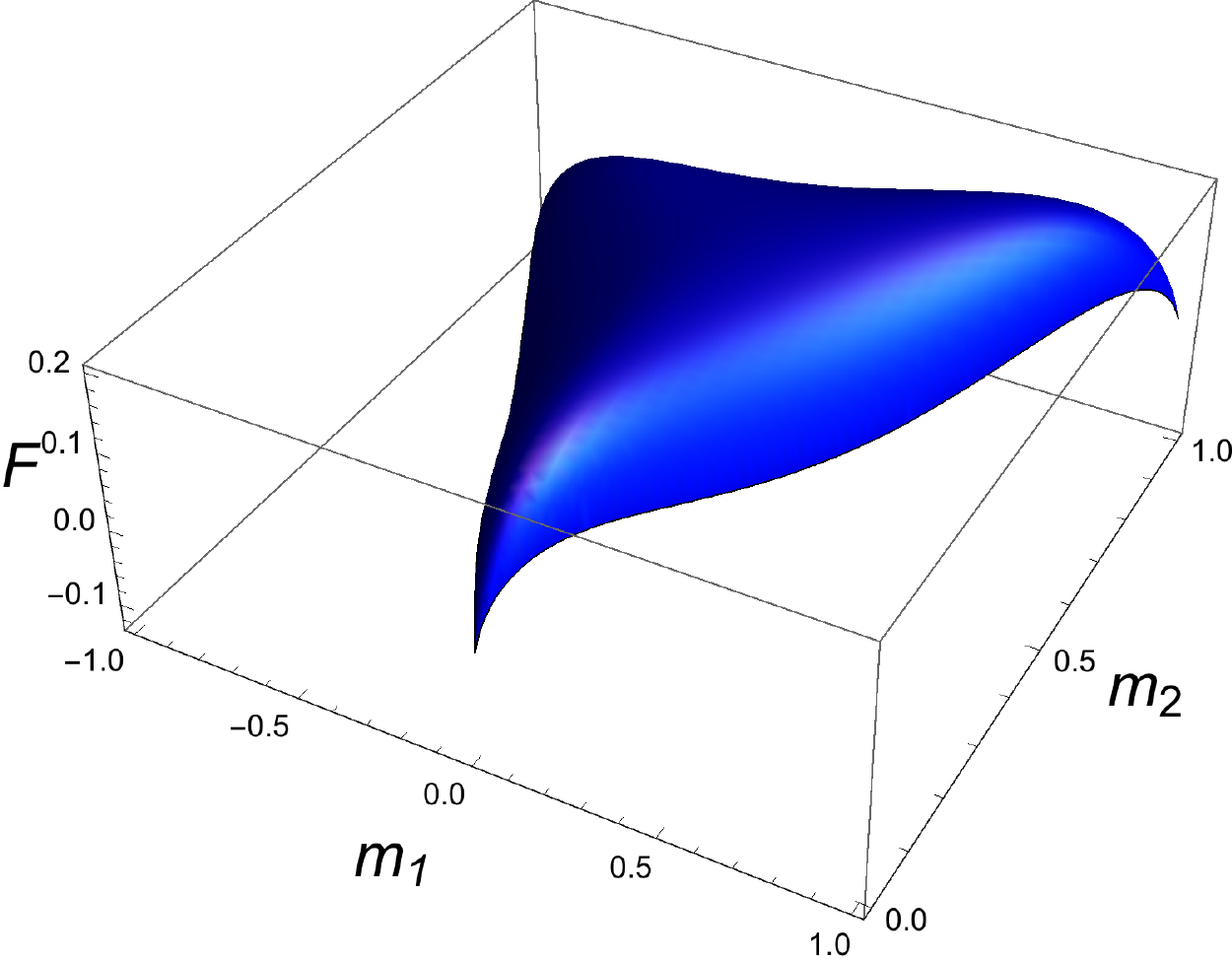}   
 \end{subfigure}
  \caption{Free energy as a function of order parameter for the choice of coupling constants $x = 0.08$, $y = -0.68$ and $t = 1.3 > t_{c}$}
\label{fig:free_energy}
\end{figure}

\begin{figure}[htb]
\centering
\begin{subfigure}[b]{0.18\textwidth}
    \includegraphics[width=\textwidth]{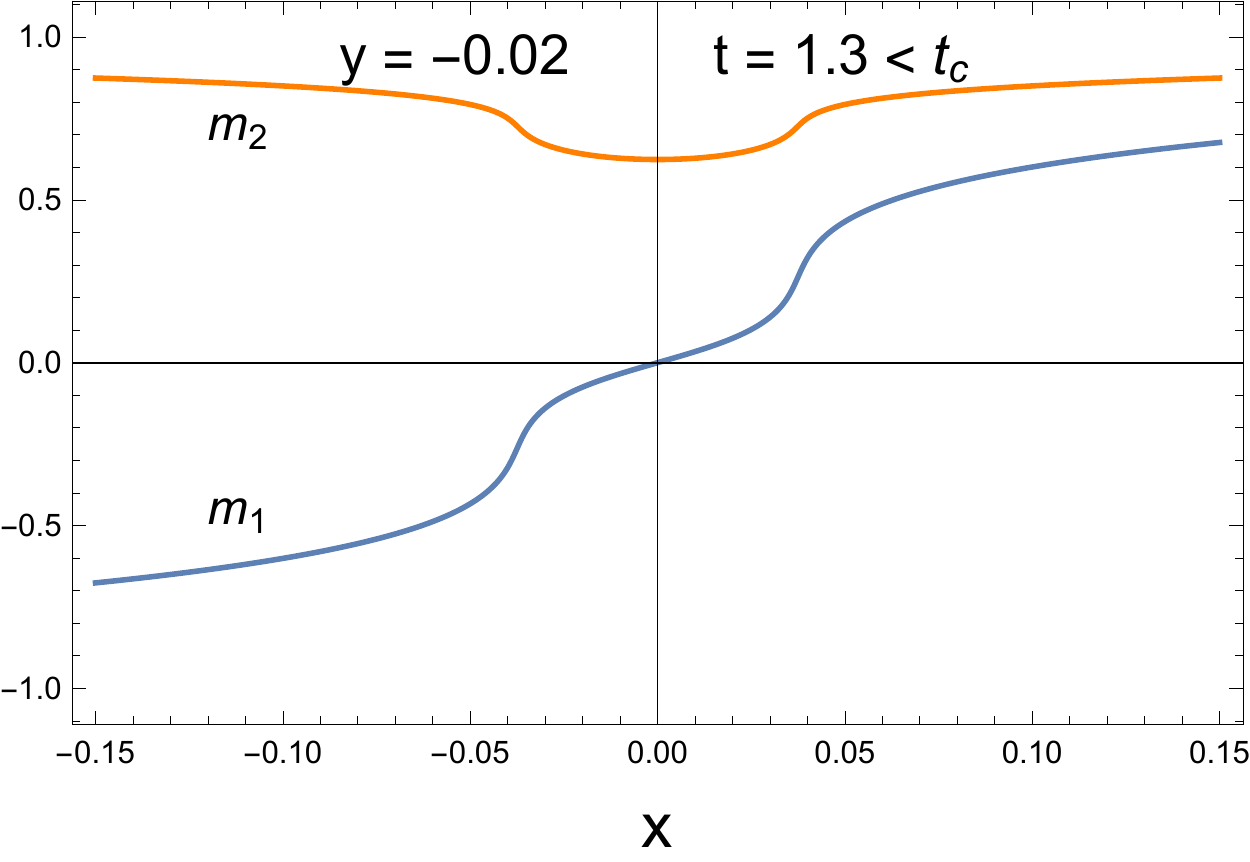}   
  \end{subfigure}
\quad
  \begin{subfigure}[b]{0.18\textwidth}
    \includegraphics[width=\textwidth]{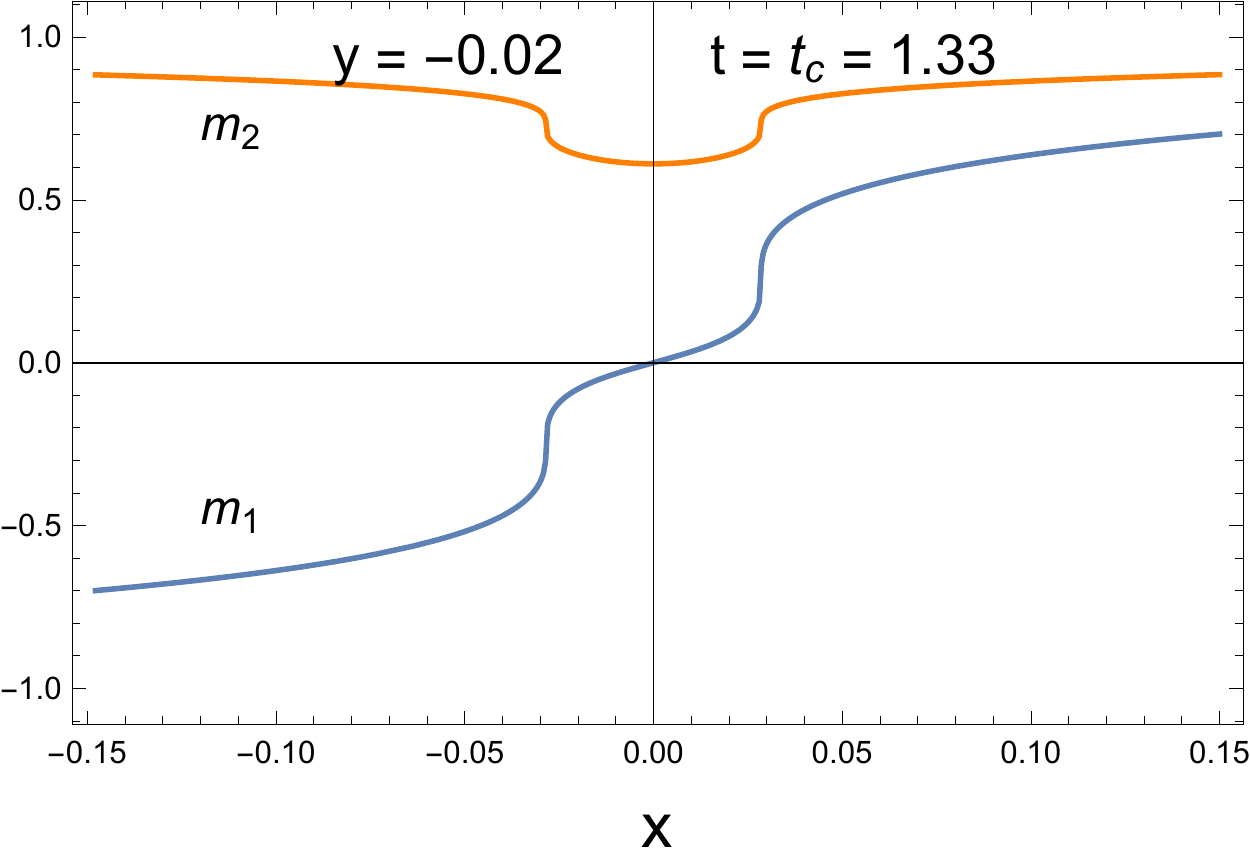}
  \end{subfigure}
\caption{\small Gradient catastrophes of the order parameters along the line $y = - 0.0202147$. The critical time is $t_{c} = 1.3263$.}
  \label{fig:critprofile2}
\end{figure}

Figures~\ref{fig:critprofile} shows the profile of order parameters for a particular choice of the field $y$ at different times. As expected in correspondence of the cusp point on the loop, at the critical ``time" $t_{c}$, both moments develop a gradient catastrophe and their profile becomes multivalued for $t > t_{c}$. Hence, the system admits multiple equilibrium states, but the physical state is selected by the maximum of the free energy function $F$.  Figure~\ref{fig:free_energy} shows the free energy as a function of the order parameters for the above choice the coupling constants. The graph clearly displays multiple maxima corresponding to equilibrium states. Figure~\ref{fig:critprofile2} shows moments gradient catastrophes on the loop. 

{\it Concluding remarks. }The dressing procedure described above arises a general and elementary approach to construct the mean field solution of  statistical mechanical models. The free energy of the model is obtained as the solution of an integrable equations of Hamilton-Jacobi type with a suitable initial condition and the equations of state are obtained as stationary points of the free energy expressed as a function of the order parameters.
   
As mentioned above, all the results obtained for $q =3$ can be extended to arbitrary $q$ and in the case of vanishing external fields they are consistent with standard results for zero external fields. In particular, the free energy of the form
\begin{equation}
\label{Fsolgen}
F=\sum_{k=1}^q p_{k}^{2} t  - \sum_{k=1}^q p_{k} \log p_{k}+\sum_{k=1}^{q-1} x_k m_{k}
\end{equation}
can be viewed as a $(q-1)$-parameter deformation of the classical formula~\cite{Mussardo},
where $p_k$ is the probability to observe a spin in the state $a_k$. 
We have that, in general, probabilities $p_k$ are linearly parameterised in terms of the moments $m_k$ as follows
\begin{equation}\label{mvsp} 
p_k=\sum_{l=1}^{q-1}c_{kl}m_l+d_{k}
\end{equation}
where coefficients $c_{kj},d_k$ are determined 
by imposing the conditions $p_k(m_1=a_j,m_2=a_j^2,...,m_{q-1}=a_j^{q-1})=\delta_{kl}$.
Introducing the $q\times q$ matrix $C$ whose first column is the vector $(d_1,...,d_q)$ and the remaining entries are $C_{i,j+1}=c_{ij}$, the condition~(\ref{mvsp}) reads as $CW(a_1,...,a_q)=I_{q}$ where $I_{q}$ is the identity matrix and  $W(a_1,...,a_q)$ is the Vandermonde matrix. Therefore, introducing the vectors ${\bf m}=(1,m_1,...,m_{q-1})$ and  ${\bf p}=(p_1,p_2,...,p_q)$ the solution to the condition~\eqref{mvsp} takes the simple and explicit form  ${\bf p}=W(a_1,...,a_q)^{-1}{\bf m}$.
\newline
\newline
{\it Acknowledgements}. Authors would like to thank: Gesualdo Delfino for drawing their attention to the Potts model, Adriano Barra, Costanza Benassi, Giovanni De  Matteis, Francesco Giglio for useful discussions and references. PL is partially supported by MIUR - FFABR funds 2017. AM is  supported by The Leverhulme Trust RPG 2017-228. Authors are also thankful to London Mathematical Society and GNFM - INdAM for supporting activities that contributed to the research reported in this paper.

\end{document}